\documentstyle[eqsecnum,aps]{revtex}

\newcommand{\be}{\begin{equation}}
\newcommand{\bea}{\begin{eqnarray}}
\newcommand{\eql}[1]{\label{eq:#1}}
\newcommand{\ee}{\end{equation}}
\newcommand{\eea}{\end{eqnarray}}
\newcommand{\ec}[1]{\ref{eq:#1}}
\newcommand{\ps}{\int {d^3k\over (2\pi)^3} }

\begin{document}
\title{Bending of Light by Vector Perturbations}
\author{S. Dodelson$^1$ and C. Wiegert$^2$}
\address{$^1$ NASA/Fermilab Astrophysics Center, P. O. Box 500, Batavia, Illinois 60510, USA}
\address{$^2$ Department of Physics, The University of Chicago, 
Chicago, IL~~60637-1433, USA}

\date{\today}
\maketitle

\begin{abstract}
A number of different mechanisms exist which can produce vector
perturbations to the metric. One might think that such perturbations
could deflect light rays from distant sources, producing observable
effects. Indeed, this is expected to be the case for scalar
perturbations. Here we show that the deflection from vector
perturbations is very small, remaining constant over large distances,
similar to the deflection due to tensor perturbations (gravity waves).
\vskip 0.2in 
\end{abstract}

\section{Introduction}

Light travels along geodesics. As such its path is determined by the
gravitational metric. When the metric deviates from a simple Minkowski
form, light ceases to travel in straight lines. Were we to detect
deviations from straight-line travel, we would be probing the very
structure of the gravitational metric.  In many cases, particularly in
cosmology, knowledge of the metric is invaluable. For example, if we
detected scalar perturbations to the metric, we would learn something
about the source of these perturbations, the mass distribution.  Since
we are otherwise limited to information about the distribution of luminous
objects such as stars and galaxies, direct information about the mass is
tremendously important.  There have been numerous successful efforts
along these lines recently, among them maps of the distribution of mass
in clusters of galaxies \cite{SQUIRES}\ and detections of massive 
compact objects in the
halo of our galaxy\cite{MACHO}. Future efforts will likely 
reveal much about the
mass distribution in the universe\cite{WEAK}, offering independent estimates of
quantities such as the power spectrum.

Over the years a number of groups \cite{HISTORY,KJ} have explored the
possibility of detecting tensor perturbations to the metric in this
fashion.  Among other reasons, tensor perturbations are interesting 
because they are produced
during inflation\cite{ABBOTT}. Thus a direct probe of tensor
perturbations in principle gives us information about inflation.
However, the consensus is now that such perturbations are much harder to
detect than are scalar perturbations.  
A scalar perturbation to the metric can coherently add to the
displacement of light over very large distances. This is not true for
tensor perturbations.

These studies naturally lead to the question of whether vector
perturbations to the metric can be detected. This is more than an
academic question: Recent studies have shown that defect theories of
structure formation produce large vector perturbations to the
metric\cite{ALLEN,PST}.  Measuring light deflection via vector
perturbations is one way then to search for elusive topological
defects. Other, less speculative, ways of producing vector perturbations
include magnetic fields which should excite vector modes.

Here we study the effect of vector perturbations on light propagation.
In a random field, the deflection due to scalar perturbations grows as
$D^{3/2}$ where $D$ is the distance travelled. The deflection due to
tensor modes was shown \cite{KJ} to grow only logarithmically as $D$
gets large. For vector modes, we find even less of an effect; the rms
deflection is constant over large distances. Thus we expect light
deflection to be an inefficient way to search for vector perturbations.
Section II presents a handwaving summary of the Kaiser-Jaffe argument
for why tensor modes do not produce deviations and extends this argument
to vector perturbations. Section III makes this argument more rigorous.

\section{Light Deflection in the Presence of Plane Wave Metric
Perturbation}

We write the metric as
\be
g_{\mu\nu} = \eta_{\mu\nu} + h_{\mu\nu}
\eql{METRIC}
\ee
where the Minkowski metric $\eta_{\mu\nu} = (1,-1,-1,-1)$. 
We will use conventions in which Greek indices run over
all four space-time coordinates and Latin indices are
spatial; also we set $c=1$. 

Consider light travelling in the $\hat z$ direction in the
presence of a single plane wave $h_{\mu\nu}$ with wavevector
\be
\vec k = k\left( \sqrt{1-\mu^2}, 0, \mu\right) .
\eql{DEFK}\ee 
Scalar perturbations are standing waves so 
$h_{\mu\nu}\propto e^{i\vec k\cdot \vec x}$,
while tensor perturbations travel at the speed
of light so $h_{\mu\nu}\propto e^{i(k t - \vec k\cdot \vec x)}$. Kaiser
and Jaffe showed that the displacements in the directions perpendicular
to the direction of propagation ($\hat z$) obey the geodesic equations
\bea
\ddot d_{\rm Scalar} &\propto& k e^{i k z \mu} \eql{GEOSCA}
\\
\ddot d_{\rm Tensor} &\propto& k (1-\mu)^{3/2} e^{i k z (1-\mu)} .
\eql{GEOTEN}
\eea
Here dots denote derivative with respect to position $z$
and
the perpendicular displacement is $d$. 
The change in the direction, or the displacement angle, after the photons have
travelled a distance $D$ is therefore
\bea
\dot d_{\rm Scalar} &\propto& {1\over \mu} \left[  e^{i k D \mu} - 1\right] \eql{DEFLECTSCA}
\\
\dot d_{\rm Tensor} &\propto&  (1-\mu)^{1/2}\left[ e^{i k D (1-\mu)} - 1\right].
\eql{DEFLECTTEN}
\eea
In the scalar case, $\vec k$-modes perpendicular to the $\hat z$ direction in
which light is travelling (i.e. those with $\mu=0$) produce an
abnormally large displacement angle. Expanding out the exponential, we
find $\dot d_{\rm Scalar} \sim kD$ as long as $\mu < 1/(kD)$. As the
light travels further and further, its displacement angle gets larger
and larger. This is physically reasonable in this simple case where the
metric consists of only one plane wave. For, the photon can indeed get a
large kick by simply travelling perpendicular to the direction in which
the field is changing. It then experiences a constant force, getting a
constant kick and corresponding boost in the displacement
angle. For
tensors the situation is completely different. In that case, for the
light to see a constant field, it needs to travel along with the gravity
wave. That is, to experience a coherent push, the photon needs to travel
in the direction along which the field is changing, $\mu=1$.  Due to the
$(1-\mu)^{1/2}$ factor in front, though, the displacement when
travelling in this direction is zero. So the typical displacement angle
of light travelling in a tensor perturbation will be of order the field
strength.  It will {\it not} be enhanced by a factor of order $kD$ as it
travels a long distance.  Tensor perturbations do not produce observable
light deflections because distortions are suppressed when light travels
in the resonant direction.

How does light behave in the presence of vector modes? Consider the following
vector field
\be
h_{\mu\nu} = \left(
	\matrix{0 & 0 & 0 & 0 \cr
		0 & 0&  -\sqrt{1-\mu^2} & 0 \cr
		0 & -\sqrt{1-\mu^2} & 0 & - \mu \cr
		0 & 0 & - \mu& 0}
	\right) e^{i\vec k\cdot \vec x},
\ee 
where $\vec k$ is again defined as in \ec{DEFK}. Borrowing from the next section, we
write down the geodesic equation for light travelling in the presence of this 
metric:
\be
\ddot d^i_{\rm Vector} = -{\partial h_{iz} \over \partial z} 
= \delta_{iy} i k \mu^2  e^{ik z \mu} .
\eql{GEOVEC}
\ee
Comparing \ec{GEOVEC} with \ec{GEOSCA}, we see that the change in
the deflection is suppressed
by a relative factor $\mu^2$ in the resonant direction. Thus,
even if $k z\mu$ is small, the deflection is still small;
there is no resonance. All modes contribute
an equal (small) amount; there is nothing special about the $\mu=0$ mode.
There will be no accumulated displacement as light travels long distances.

\section{Deflection in Random Vector Field}
 
We now make the argument of the previous section more rigorous,
introducing a general vector perturbation as a random sum 
over plane waves, and solving the geodesic
equation for a light ray. 

In synchronous gauge, a general vector perturbation has
only space-space components ($h^{0i} = h^{00} = 0$). If
we Fourier transform $h$, the spatial components are
\be
h_{ij}(\vec x) = \ps e^{i\vec k \cdot \vec x}
\left[ h_1(\vec k) \bigg( \hat m_{1i} \hat k_j + \hat k_i \hat m_{1j} \bigg)
+ i h_2(\vec k) \bigg( \hat m_{2i} \hat k_j + \hat k_i \hat m_{2j} \bigg)
\right]
\eql{DEFHIJ}
\ee
where $\hat m_1$ and $\hat m_2$ are  two unit vectors orthogonal
to wave direction $\hat k$ and to each other. The factor of $i$
is inserted here to ensure that reality implies $h_a(\vec k)
= h^*_a(-\vec k)$ for $a=1,2$.
With no loss of generality, we can choose the propagation
direction of light to be in the $\hat z$
direction. Then, write the three orthogonal vectors as
\bea
\hat k &=& \left(\sqrt{1-\mu^2}\cos\phi,\sqrt{1-\mu^2}\sin\phi,
		\mu \right)
\\
\hat m_1 &=& \left( \sin\phi , -\cos\phi, 0 \right)
\\
\hat m_2 &=& \left( -\mu\cos\phi , -\mu\sin\phi , \sqrt{1-\mu^2} \right).
\eql{DEFORTHO}
\eea
Note that we have assumed here that the vector field is time independent.
We argue that this is a conservative assumption. In a cosmological setting,
vectors fields decay over time; the only way they can be important is if
they are continuously seeded and so remain relatively constant with time.

Consider a photon travelling through this field with direction
$\hat z + \dot{\vec d}$, where 
the dot denotes $d/dt = \hat n^\mu\partial/\partial x^\mu$
and the zero order direction is $n^\mu = (1,\hat z) = (1,0,0,1)$.
Since the vector field is assumed to be time independent,
$d/dt = \partial/\partial z$. 
The geodesic equation for the light is then
\bea
\ddot d^i &=& -\Gamma^i_{\mu\nu} \hat n^\mu \hat n^\nu = 
-\eta^{ij} \hat n^\mu \hat n^\nu\left[  {\partial h_{j\mu} \over \partial x^\nu}
- {1\over 2} {\partial h_{\mu\nu} \over \partial x^j} \right]
\nonumber\\
&=&
- \left[ {\partial h_{iz} \over \partial z}
- {1\over 2}  {\partial h_{zz} \over \partial x^i} \right]
\eql{GEODESIC}
\eea
The last equality follows since
$h_{\mu\nu}$ has no time components.
Upon inserting our expression for $h_{\mu\nu}$ into \ec{GEODESIC}\ we
find
\be
\ddot d^i = \dot H^i
\ee
with 
\be
H^i \equiv  \ps  \mu e^{i k z \mu}
\left( -h_1\sin\phi + i h_2\mu\cos\phi, h_1\cos\phi + i h_2\mu\sin\phi\right).
\ee
Since the displacement angle $\dot d^i \propto H^i \propto \mu$, 
it does {\it not} get a large contribution from modes with
$\mu=0$; in fact these contribute relatively little. Rather, $\dot d$ is
of order the field strength $h_1,h_2$.

If $h_1(\vec k)$ and $h_2(\vec k)$ are random fields, then the 
change in the direction of a photon travelling from $z=0$
to $z=D$ will be zero on average. The mean square direction change can
be calculated:
\bea
\langle | \delta \dot{\vec d} |^2\rangle 
&=&
\langle ( \dot d^i(D) - \dot d^i(0) )  ( \dot d^i(D) - \dot d^i(0) ) \rangle
\nonumber\\
&=&
{1\over 2\pi^2} \int_0^\infty dk k^2 \int_{-1}^1 d\mu 
\left( \mu^2 P_1(k) + \mu^4 P_2(k) \right) \left( 1 - \cos(k\mu D) \right)
\eql{DIRCH}
\eea
where the power spectra are defined so that
\be
\langle h_a(\vec k) h_b^\ast(\vec k')\rangle = 
\langle h_a(\vec k) h_b(-\vec k')\rangle = 
(2\pi)^3 \delta(\vec k - \vec k') \delta_{ab} P_a(k) .
\ee
The oscillatory term in \ec{DIRCH}\ is irrelevant for large
distances, and we are left with
\be
\langle |\delta \dot{\vec d} |^2\rangle =
{2\over 3} \langle {h_1}^2 \rangle + 
{2\over 5} \langle {h_2}^2 \rangle
\ee
So, even after travelling large distances in a vector field,
light has experienced little directional change. This is identical
to the case of a tensor field but dramatically different than the
scalar field, for which $\langle |\delta \dot{\vec d} |^2\rangle
\propto D$.

Kaiser and Jaffe showed that many observables are governed by
the power spectrum
\be
P_H(k) \equiv \int_{-\infty}^{\infty} dz e^{-ikz} \langle \vec H(0)
\cdot \vec H(z) \rangle
.\ee
Typically the mean square displacement after a distance $k^{-1}$
in a random field is given by $P_H/k$ which is proportional to
$k^{-3}$ for scalar perturbations but $k^0$ for tensor
perturbations.

We now calculate this power spectrum for vector modes:
\bea
P_H(k) & = & 
{1\over 4\pi^2} \int_{-\infty}^{\infty} dz \int_0^\infty dk' 
\int_{-1}^{1} d\mu
{k'}^2 \left( \mu^2 P_1(k') + \mu^4 P_2(k') \right) e^{iz(k'\mu - k)}
\nonumber \\
& = &
{1\over 2\pi} \int_0^\infty {k'}^2 dk' \int_{-1}^1 d\mu
\left( \mu^2 P_1(k') + \mu^4 P_2(k') \right) \delta(k'\mu - k)
\nonumber \\ 
& = &
{1\over 2\pi} \int_k^\infty k' dk' \left[ 
\left({k \over k'}\right)^2 P_1(k') + \left({k \over k'}\right)^4 P_2(k')
\right] \eql{PH}
\eea
On large scales (small $k$) then, 
$P_H(k)/k \propto k$ for vector modes. Thus in the physical
cases where tensor modes produce a logarithmic divergence, vector
modes produce no such divergence. 

One example of this is the question of the angular deflection of
an image from its unperturbed location. On average, the image is unchanged,
but the rms angular deviation is
\be
\delta \theta_{rms} = {1\over D} \left\langle \left( 
\int_{z_0}^{z_1} dz H(z) \right)^2 \right\rangle^{1/2}
,\ee
where the light starts at $z_0$ and travels a distance $D$ to
$z_1$.
Kaiser and Jaffe showed that this angular deviation is proportional
to $D^{1/2}$ for scalar perturbations and $\sqrt{\ln(D)}/D$ for
tensor perturbations. To calculate it for vector perturbations, we
again follow Kaiser and Jaffe to write
\bea
\delta \theta_{rms} &=& {1\over D} \left(
{2\over \pi} \int {dk\over k^2} P_H(k) \sin^2(kD/2)  \right)^{1/2}
\nonumber\\
&=&
{1\over D} \left(
{2\over D\pi^2} \int_0^\infty {dk'\over k'} 
\int_0^{k'D/2} dx \sin^2x [ P_1(k') + (2x/k'D)^2 P_2(k') ]
  \right)^{1/2}
,\eea
where in the last line here, we have used \ec{PH}, changed
the order of integration, and introduced the dummy variable $x=kD/2$.
Performing the $x$ integral, but keeping only terms to highest order in $k'D$
 leads to
\be
\delta \theta_{rms} = 
{1\over D} \left(
{1\over 2\pi^2} \int_0^\infty dk'
\bigg[
P_1(k') + {1\over 3} P_2(k')
\bigg]
  \right)^{1/2}. 
\ee
So the rms displacement angle is of order 
$\langle h^2\rangle^{1/2}/k_V D$ where $k_V$ is
the wavenumber where the power spectrum peaks. This is even smaller
than the corresponding displacement angle for tensor modes, which was
enhanced (slightly) by a logarithm.

\section{Conclusions}

Even if there is a background of vector modes perturbing the
gravitational metric, light should travel virtually undeflected
over large distances. This conculsion is markedly different than what
we expect if there are scalar perturbations to the metric, since
scalar perturbations can act coherently over large distances. Even in the
best case, where vector modes remain constant, any coherent action is
defused by a suppression of the perturbation in the resonant direction
(i.e. the $\mu^2$ factor in \ec{GEOVEC}). Vector modes, just like tensor
modes, do not bend light.

We thank Andrew Jaffe and Albert Stebbins
for useful discussions.
This work was supported in part by the DOE
and by NASA grant NAGW-2788 at Fermilab.

\end{document}